\documentclass[11pt,preprint,flushrt]{aastex}
\usepackage{amsmath}
\usepackage{times}
\allowdisplaybreaks[1]

\begin{document}

\title{Telescope Optics and Weak Lensing: PSF Patterns due to Low
Order Aberrations}

\author{Mike Jarvis$^1$, Paul Schechter$^2$, Bhuvnesh Jain$^1$}
\affil{
${}^1$Department of Physics and Astronomy, University of Pennsylvania,
Philadelphia, PA 19104\\
${}^2$Department of Physics, MIT, Cambridge, MA 02138\\
}

\begin{abstract}
In weak lensing investigations, galaxy shapes are corrected for the
convolution by the point spread function (PSF) using stellar images. 
In this paper we use physical models of the telescope optics to
understand the spatial variation of the
PSF in the image plane.  We introduce a set of parameters to model the
key aberrations, which include defocus, focal plane tilt, primary
and off-axis astigmatism, and coma. 
We also include the effects of guiding and seeing. We
test our model with data from the Blanco 4 meter telescope in Cerro
Tololo, Chile. We find that the physical model describes a substantial
part of the PSF size and anisotropy over the field of view (over 90
percent of it, based on a chi-squared metric). We
identify the primary contributors to the PSF patterns and study their 
covariances and principal components. We also identify correlations with
the effect of gravity on the telescope. 
Finally, we discuss the improvements in PSF estimation
that may be achieved by combining the physical model in this study
with the purely empirical approach of Jarvis and Jain (2004). 
\end{abstract}


\section{Introduction}

The weak gravitational lensing of background galaxies by foreground
galaxies has proven itself a powerful technique for studying the
largest structures in the universe \citep[e.g.][]{schneider06}.  But on the
largest angular scales, the coherent distortions of galaxy images due to
lensing can be as small as one part in one thousand.  The reliable
measurement of such small effects requires a thorough understanding of
instrumental and observational effects that might masquerade as weak lensing.

No telescope produces perfectly circular images. The extent to which
those images deviate from perfect circularity depends upon the design
of the telescope and the degree to which the telescope maintains its
alignment.  Gravity, thermal effects, mechanical oscillations and
operator error can all cause images to be elliptical.

The practioners of weak lensing calibrate their instruments by
measuring the shapes of stars, which
are effectively delta functions before passing through the atmosphere and telescope.
The typical high galactic latitude field has relatively few stars, but by
averaging over many such fields 
%(e.g. Clowe; Jarvis??) 
or by using 
fields with many more stars \citep[e.g.][]{hoekstra04}, one can 
improve the PSF estimation.

Unfortunately most of the phenonmena that cause elliptical PSF images are
time variable.  Typical exposures and time averages correct only for
that part of the instrumental PSF that is time invariant.

With too few stars in a single exposure to produce a precise map of the PSF,
a model is needed to track such temporal
variations.  Such models might be be calculated from first principles
(``theoretical'' models), purely empirical, or some combination of the
two.  

An example of the empirical approach would be the efforts of \citet{Jarvis04}
who looked at the principal components of the PSF shapes
for a thousand images taken with the BTC and Mosaic 
cameras on the Victor Blanco telescope.  For every exposure one can then
calculate the projection of its PSF pattern onto a limited subset of
those principal components and correct accordingly.
\citet{Jarvis04} also offer physical interpretations for the largest of
their principal components.  Their first principal component
seems to be telescope focus.  The second appears to reflect
tracking errors.

There are several benefits if the physics dominating one or more of these
principal components can be identified.  Foremost, one might then
address the underlying cause and remove the source of the pattern.
Second, there is no guarantee that a single principal component has a
single physical cause (and no guarantee that physical effects produce
orthogonal principal components).  If one models known causes of
the PSF shape directly, it opens the possibility that physical causes can
be identified for the remaining principal components.  Finally a
physical model may be more accurate than a principal component, which
is derived from noisy and incomplete data.

There are three main sources of PSF ellipticity produced by a
telescope (apart from those that are a direct consequence of the
telescope design): guiding errors, misaligned optics, and deformations
of the primary mirror. These sources can vary with time, hence their
effect varies from one exposure to another.  
The first and third of these however have the same
effect across the entire field, producing the same shape in all
star and galaxy images in a given exposure. This makes it easy to
correct for them. 

Misaligned optics, however, produce aberration patterns that vary
across the field.  These patterns take particularly simple forms when
expanded as polynomials in wavefront error \citep[e.g.][]{mahajan91,
schroeder99}. 
The lowest order terms relevant for weak lensing are called focus,
astigmatism and coma.  As described in \S2 below, nine numbers
suffice to characterize the focus, astigmatism, and coma patterns that
result from first order telescope misalignments.  The problem of
modeling the PSF due to optical aberrations may
therefore not be as daunting as it might first seem.

In \S2 we review low order telescope aberrations.  In \S3 we describe
the patterns produced by telescope misalignments.  In \S4 we discuss
the aberrations produced by deformations of the primary mirror.  In
\S5 we examine the PSF patterns from a few hundred exposures taken
with the Blanco telescope. We interpret them in the context of
telescope misalignments and primary mirror deformations.

\section{Review of Low Order Aberrations}

Image aberrations are conveniently described using a power series
expansion of the wavefront, with cordinates $\rho$ and $\theta$
described on the telescope pupil (i.e. the plane of the primary mirror), 
where the dimensionless
radial coordinate $\rho$ is taken to be unity
at the outer boundary of a circular pupil.  \citet{mahajan91} and \citet{schroeder99}
give, respectively, exhaustive treatments of aberration theory
in general and telescope aberrations in particular. The two lowest
order polynomials describe a change in phase of the wavefront and a
tilt of the wavefront, neither of which is relevant for the present
discussion.  The next lowest terms in the expansion are called defocus,
astigmatism, coma and spherical aberration, all of which are relevant
to the present discussion.

One can look  at the PSF either in the focal (image) plane or the
wavefront projected onto the pupil.  In physical optics they
contsitute a Fourier pair.  For weak lensing, the image plane PSF is
the primary quantity of interest, but telescope aberrations
are more easily described in terms of their effect on  the wavefront
in the pupil plane.  This is because, in the pupil plane, the effect of multiple
aberrations add linearly in the wavefront, but have a more complex
interaction in the image plane.  That is the approach followed in this section. In
\S3 we will describe the variation of the aberrations in the image
plane. 

\subsection{Defocus}

The focus is adjusted by moving the focal plane along the optical axis
(or by moving a mirror to create the same effect).  However, we can 
project the effect of the defocus on the wavefront back to the primary mirror
to describe it in pupil coordinates.  

The resulting defocus,
$\delta \lambda _{\rm defocus}$, is given by
\begin{equation}
\delta \lambda _{\rm defocus}  = A_{\rm defocus} \rho^2
\end{equation}
Assuming, as is the case for ground based telescopes, that the images
are not diffraction limited, the effect of the wavefront on image
quality is readily understood by considering the gradient of the
wavefront, which is, 
\begin{equation}
\nabla \delta \lambda _{\rm defocus} = 2A_{\rm defocus}
\left( \rho \cos \theta \hat x + \rho \sin \theta \hat y \right)
\label{del_defocus}
\end{equation}.

\begin{figure}[t]
\epsscale{1}
\plotone{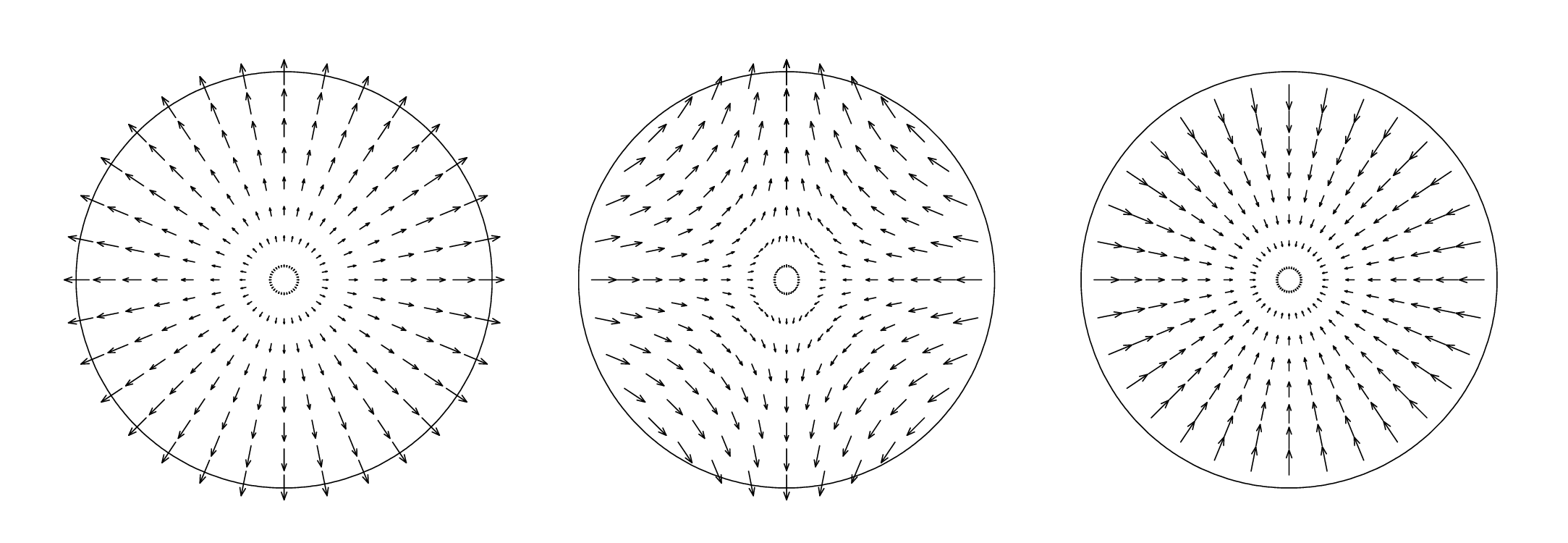}
\caption[]{\small 
Wavefront slope errors projected onto a circular pupil.  The left
and and right show slope errors for positive and negative defocus.
The center shows slope errors for pure astigmatism.  At every point
in the three figures the slope errors have the same magnitude.  But
the slope vectors for astigmatism are reflected about the vertical
(horizontal) axis with respect to the defocus vectors on the left
(right). Combining the astigmatism vectors with the defocus vectors
on the left (right) produces a PSF that is a stright vertical
(horizontal) line.
}
\label{focusvec}
\end{figure}

The outer panels of Figure~\ref{focusvec} shows the gradient of the wavefront for
two images with equal and opposite amounts of defocus.  It may be
helpful to think of the defocus as arising from slope errors on the
primary mirror.  Dividing the pupil into small equal area segments,
the point spread function (PSF) is then computed by plotting one point for
each segment, with the angular deflection from perfect focus
proportional to the gradient of the wavefront (the effective focal
length being the constant of proportionality).  For the two cases
illustrated in the left and right panel of Figure~\ref{focusvec}, the resulting PSF
is a boring, uniformly illuminated circular disk.

%fig 1

\begin{figure}[t]
\epsscale{1.0}
\plotone{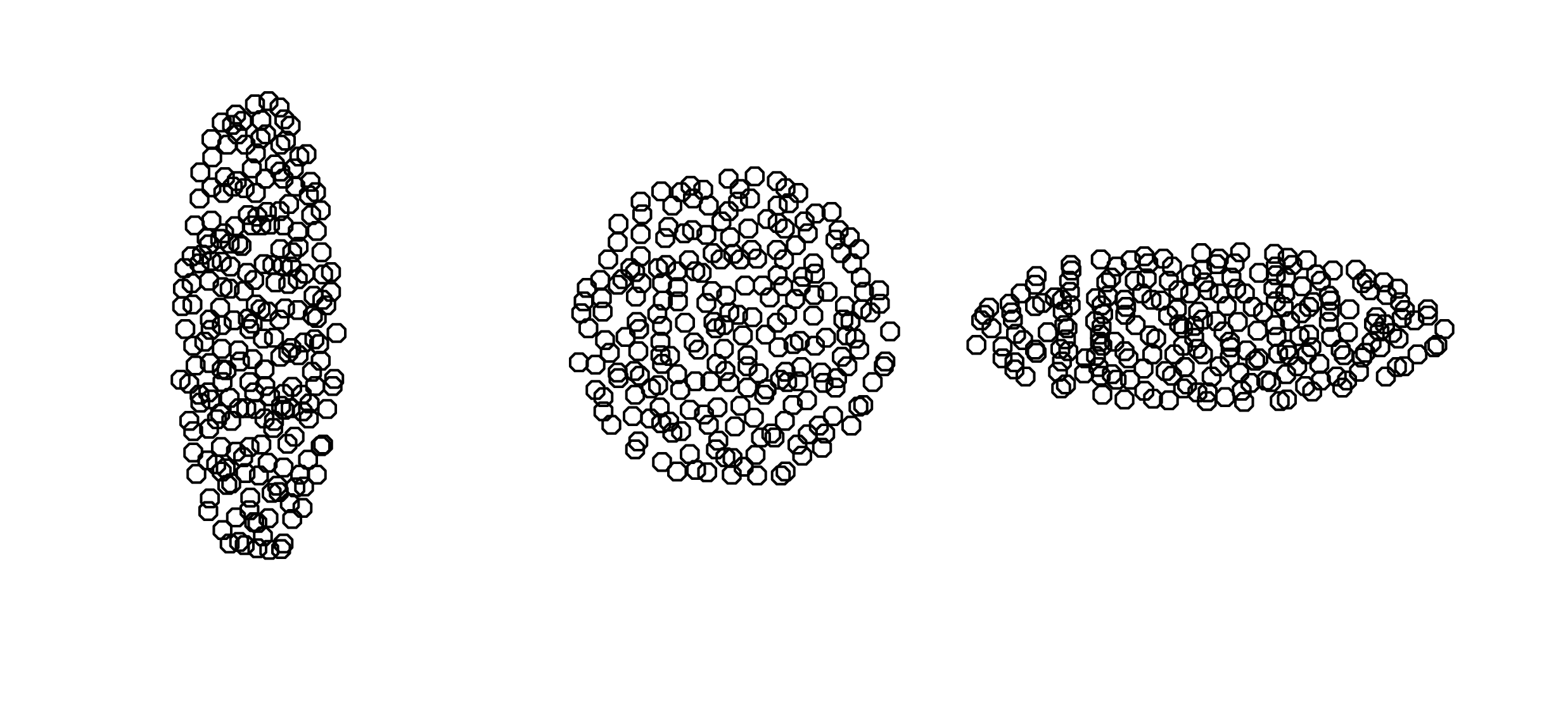}
\caption[]{\small 
The center shows the point spread function obtained from the
purely astigmatic slope errors shown in the center of Figure\ref{focusvec}.  The
left (right) shows the PSF obtained from adding one half the defocus
shown on the left (right) of Figure\ref{focusvec} to the astigmatism shown in the
center of that figure.  The spot displacements have been randomized
by roughly 10\%.  
A combination of astigmatism {\it and} defocus is
	       needed to produce elliptical images. 
}
\label{astigpm}
\end{figure}%fig 2

\begin{figure}[t]
\epsscale{0.3}
\plotone{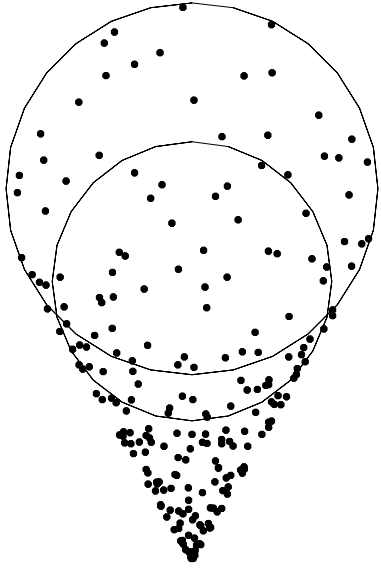}
\caption[]{\small
The point spread function produced by a purely comatic wavefront.
The larger circle indicates the locus of spots produced by the outer edge
of the pupil.  The smaller circle indicates the locus of spots
produced by points on the pupil 6/7th of the way to the edge. 
}
\label{comaspot}% fig 3
\end{figure}

\begin{figure}[t]
\epsscale{0.5}
\plotone{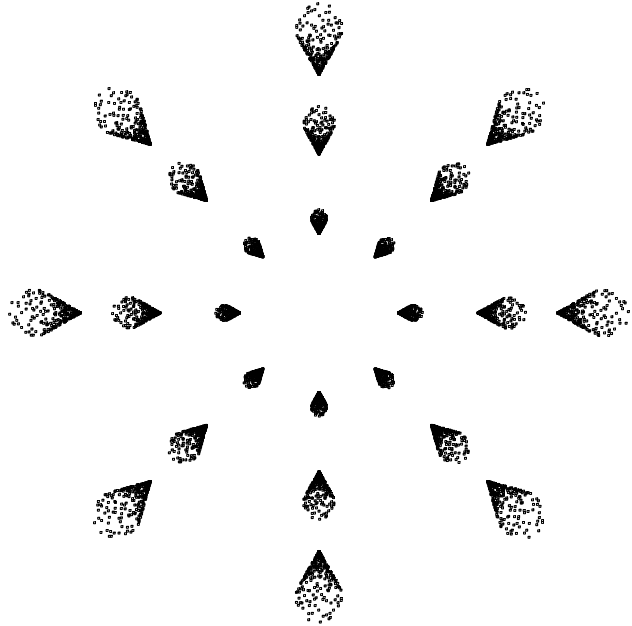}
\caption[]{\small 
Off-axis comatic point spread functions produced at the prime
focus of a parabolic mirror.  The size of the PSF increases linearly
with distance from the center of the field, and points outward.
}
\label{oaxcoma}% fig 4
\end{figure}

\begin{figure}[t]
\epsscale{0.5}
\plotone{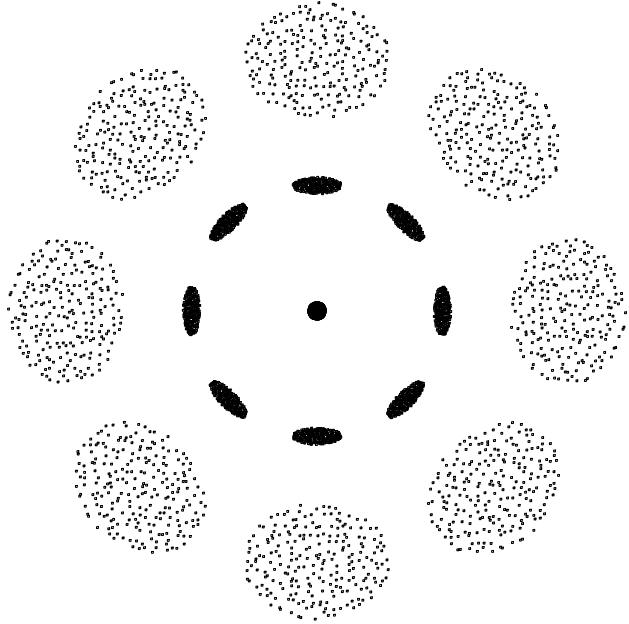}
\caption[]{\small 
Off-axis astigmatic point spread functions produced at the prime
focus of a parabolic mirror.  In the absence of defocus the PSF
would be circular and would increase quadratically with distance from
the center of the field.  A small amount of defocus, indicated by
the spot at the center, has been added to highlight the potential
for astigmatism to produce elliptical images.
}
\label{oaxastig}% fig 5
\end{figure}

\begin{figure}[t]
\epsscale{0.9}
\plottwo{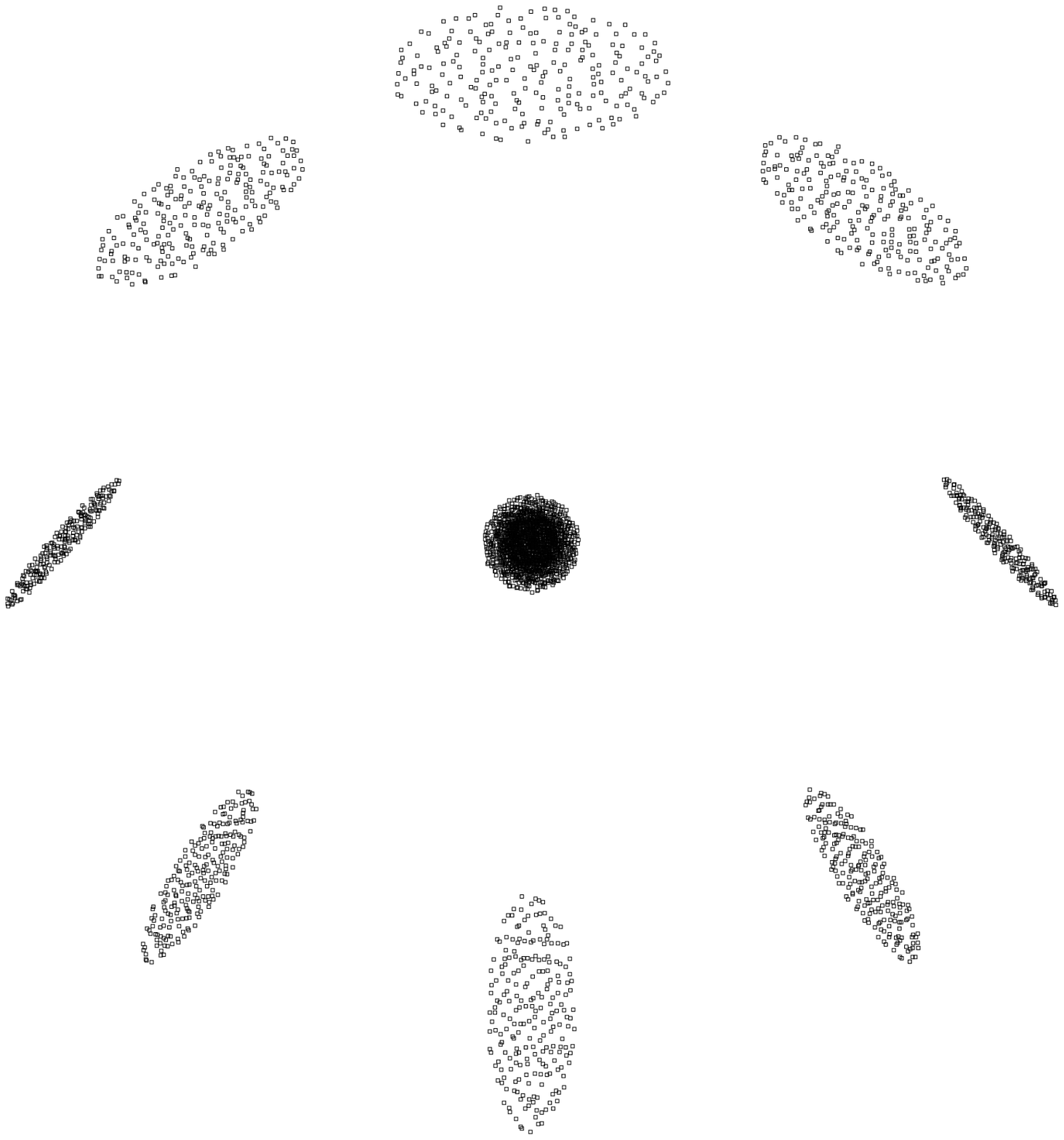}{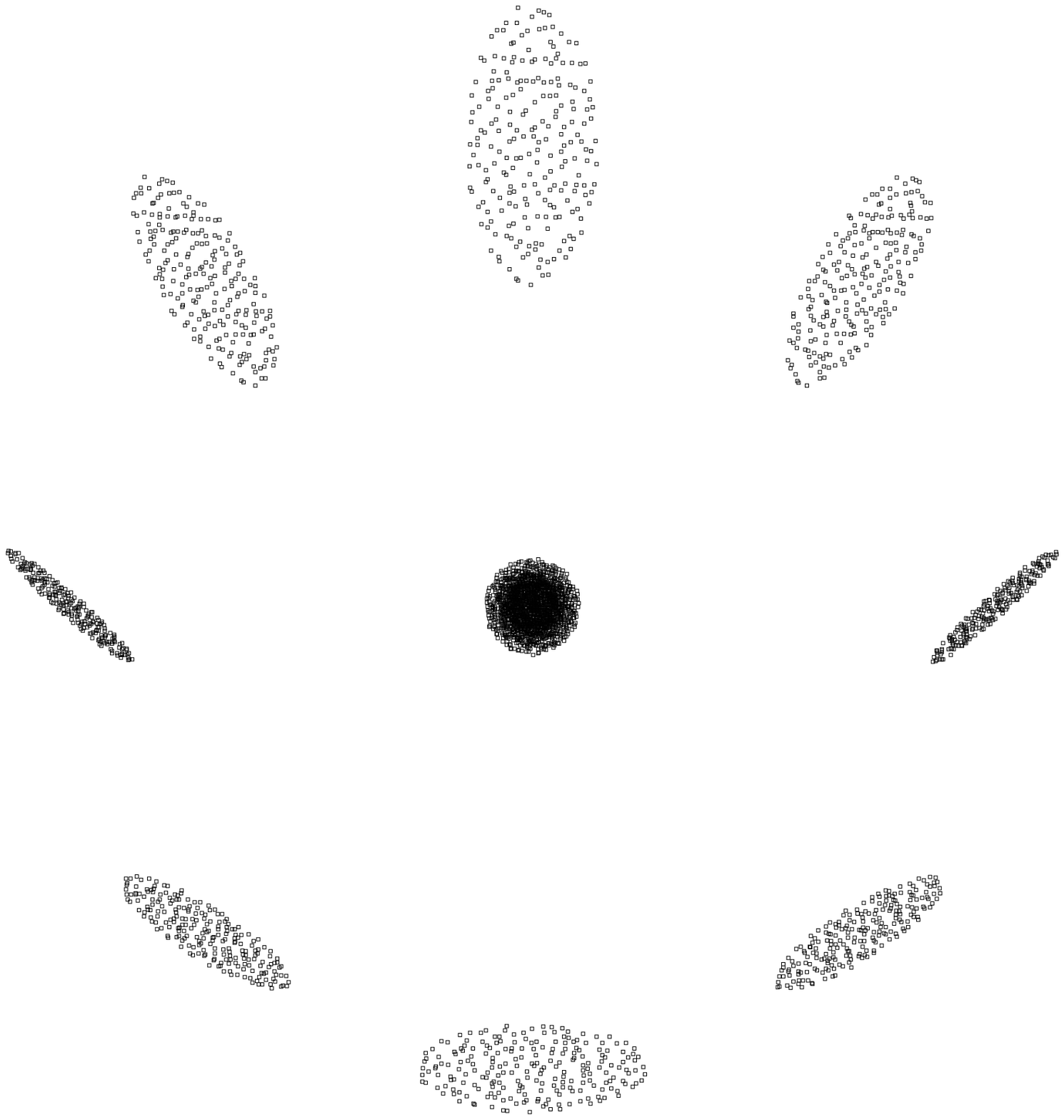}
\caption[]{\small 
The point spread function caused by miscorrecting for
off-axis astigmatism.  The figures on the left and right are obtained
by taking two copies of figure 5, shifting one vertically with
respect to the other and then subtracting one from the other.  A
small amount of defocus has been added to the figure on the left and
subtracted from the figure on the right.  If one had a misaligned
corrector, one would expect to see one pattern or the other as the
focus varied.  Since {\it all} correctors are misaligned, the
question is only one of degree. 
}
\label{badastig}% fig 6
\end{figure}

\subsection{Astigmatism}

The primary cause of astigmatism is an overall warping of the primary mirror, but as we
discuss below, there are other causes which lead to a variable astigmatism over the 
field of view.

There are two independent
components to the astigmatism of a wavefront, $\delta \lambda _{astig-c}$, and
$\delta \lambda _{\rm astig-s}$, given by
\begin{align}
\delta \lambda _{\rm astig-c}  &= A_{\rm astig-c} \rho^2 \cos 2 \theta \\
\delta \lambda _{\rm astig-s}  &= A_{\rm astig-s} \rho^2 \sin 2 \theta
\end{align}
The gradients of these two components are given by
\begin{align}
\nabla \delta \lambda _{\rm astig-c} &= 2A_{\rm astig-c}
\left( \rho \cos \theta \hat x - \rho \sin \theta \hat y \right) 
\label{del_astig1}
\\
\nabla \delta \lambda _{\rm astig-s} &= 2A_{\rm astig-s} \left(\rho \sin \theta \hat
x + \rho \cos \theta \hat y \right)
\label{del_astig2}
\end{align}

Since both components are purely quadratic in wavefront, the sum of
two purely astigmatic contributions to the wavefront is again purely
astigmatic.  One can always find a coordinate system in which the
second component is zero, so in what follows we will use a single coefficient
$A_{\rm astig}$.  The central panel of Figure~\ref{focusvec} shows the gradient of the
wavefront for the first of the two astigmatism components. (A gradient
plot for the second component would be rotated by $45\deg$.)

It is instructive to compare the lengths and orientations of the
gradient vectors in the three panels of Figure~\ref{focusvec}.  At every point in
each of the three panels the vectors are the same length.  In the
first and last panels (the two defocused wavefronts) the vectors point
in opposite directions.  In the central panel (the astigmatic
wavefront) one of the Cartesian components is the same as in the
adjacent panel while the other has the opposite sign.  This has
important consequences for the ellipticity of images and for weak lensing.

The gradient vectors in the central panel in Figure~\ref{focusvec} may be used
to calculate the PSF for the astigmatic wavefront.  This is done
simply by collecting the heads of all the vectors at the center so 
that the arrows of the vectors represent the PSF. The result is shown
in the central panel of Figure~\ref{astigpm}.  Some readers may be surprised
to see that the PSF of an astigmatic image is circular.  This is
a direct consequence of the fact that the wavefront gradient vectors are
the same as for a defocussed image, except for a switch in sign
of one of the two coordinates.  

But astigmatic images are never perfectly in focus.  If we add defocus
with plus or minus half the amplitude of the astigmatism, we get the
flattened PSFs seen in the left and right panels of Figure~\ref{astigpm}.  
Non-round PSF images are the result of astigmatism {\it and} defocus.  A
straightforward calculation shows that the difference in second
moments of such an image is proportional to the {\it product} of the
astigmatism and the defocus.  If we added and subtracted defocus with
the same amplitude as the astigmatism in Figure~\ref{astigpm}, the PSFs would be
lines of finite length.  Adding and subtracting still more defocus
produces rounder images (but with the difference between the second
moments nonetheless increasing).

%The ellipticities used by practioners of weak lensing are closely
%related to quadrupole moments.  The quadrupole moment,
%$I$, of an image with astigmatism $A_{\rm astig}$ and defocus $A_{\rm defocus}$
%is given by 
%\begin {equation}
%I \sim \frac{(A_{\rm defocus} + A_{\rm astig})^2}
% {(A_{\rm defocus} - A_{\rm astig})^2} \quad .
%\end {equation}

Astigmatic images have the character that the rays are converging too
quickly in one direction and too slowly in the orthogonal direction.
On one side of focus they converge to a line.  As one approaches focus
they start to diverge while the rays in the orthogonal direction are
still converging.  On the other side of focus these rays then converge
to a line, perpendicular to the first.

\subsection{Coma}

Coma is famously associated with Newtonian telescopes, which use a parabola for
the primary mirror.  The center of the field of view focuses correctly, but as an
image moves further from the optical axis, the parabola no longer focuses the light from
different parts of the mirror onto the same point.  Modern telescopes use corrector lenses
to account for this effect.

As with astigmatism, there are two independent
components to the coma of a wavefront, $\delta \lambda _{\rm coma-c}$, and
$\delta \lambda _{\rm coma-s}$, given by
\begin{align}
\delta \lambda _{\rm coma-c}  &= A_{\rm coma-c} \left(\rho^3 - \rho\right) \cos \theta \\
\delta \lambda _{\rm coma-s}  &= A_{\rm coma-s} \left(\rho^3 - \rho\right) \sin  \theta
\end{align}
The gradients of these two components are given by
\begin{align}
\nabla \delta \lambda _{\rm coma-c} 
&= A_{\rm coma-c} \left( \left(2 \rho^2 - 1 +\rho^2 \cos 2\theta\right) \hat x + 
\rho^2 \sin 2\theta \hat y \right) 
\label{del_coma1}
\\
\nabla \delta \lambda _{\rm coma-s}  
&= A_{\rm coma-s} \left( \rho^2 \sin 2\theta \hat x + \left(2\rho^2-1
- \rho^2 \cos 2\theta\right) \hat y \right)
\label{del_coma2}
\end{align}
We note that the $\rho$ term needs to be included so that the centroid in the image plane is zero. 
In general, coma does move the centroid of the PSF, but we care more about the shape than
the position, so we choose this form to avoid a spurious 
contribution to the second moments from a non-zero centroid.

A comatic PSF is illustrated in Figure~\ref{comaspot}.  The rays coming from the
center of the pupil lie very close together.  Those from the outer
boundary of the pupil lie on a circle offset to one side of the more
central rays.

A comatic PSF is completely specified by an amplitude and a
direction.  In this regard, the comatic PSF behaves like a vector.
The sum of two different comatic wavefronts again gives a comatic
wavefront.  Moreover, the resulting PSF has an amplitude and direction
given by the rules of vector addition.  We can therefore treat the
comatic PSF as a vector.  

By contrast, the PSF of an astigmatic image cannot be treated as a
vector.  There is a twofold ambiguity in assigning a direction to
an astigmatic PSF, hence the PSF that results from the 
addition of two astigmatic wavefronts cannot be obtained by simple
vector addition of the two astigmatic point spread functions.

\subsection{Spherical Aberration}

The last aberration we consider is spherical aberration, which, as the name suggests, is 
associated with the primary mirror being more spherical than parabolic.  
Modern telescopes with multiple optical elements can reduce both this and coma at once
to a large extent, but they can never completely eliminate both.
Therefore, while we don't expect this to be something
that varies from one exposure to the next, there may be a constant spherical aberration
due to incomplete correction in the telescope design or errors in the positions of some 
of the optical elements.

The effect of spherical aberration is given by
\begin{align}
\delta \lambda _{\rm spherical}  &= A_{\rm sph} \rho^4 \\
\nabla \delta \lambda _{\rm sph}
&= 4 A_{\rm sph} \left( \rho^3 \cos \theta \hat x + 
\rho^3 \sin \theta \hat y \right) 
\label{del_sph}
\end{align}

\section{Miscorrected Off-Axis Aberrations}

A paraboloidal mirror produces perfect images on-axis.  But off-axis
images suffer from coma, which increases linearly with distance from the
axis, and astigmatism, which increases quadratically with distance
from the axis. In this section we will consider the PSF resulting from
these two aberrations
and describe their variation in the image plane (in contrast to the
pupil plane description of wavefront error in \S2). 

Weak lensing requires wide fields, and the telescopes used to study
weak lensing have what are in effect ``correctors'' for off-axis coma
and astigmatism.  A prime focus telescope like the Blanco 4-meter at
Cerro Tololo has a multi-element transmitting corrector close to the
prime focus.  A Ritchey-Chretien telescope \citep[e.g.][]{schroeder99} like
the ESO 2.2-meter on La Silla is corrected for off-axis coma by making
the primary mirror slightly hyperbolic and balancing this with a
hyperbolic secondary that is slightly different from what it would
otherwise be in a straight Cassegrain configuration.  A transmitting
corrector close to the secondary focus (e.g. Bowen and Vaughn 1972)
then corrects for astigmatism.

For our purposes it suffices to imagine a single corrector close to
the focal plane.  If all of the optical elements are aligned the telescope
produces perfect images (or more likely very slightly imperfect images).
But if the corrector is displaced perpendicular to the optical axis,
the wrong correction is applied to the images.  The wavefront
arriving at the detector is given by the difference between the
correction needed and the correction applied.

No telescope is ever perfectly aligned.  The most careful alignment
process will still produce small but hopefully tolerable misalignments.
More seriously, thermal stresses, variable gravitational forces and human
error all lead to misalignments.

\subsection{Off-Axis Coma}
\label{off-axis_coma}

By virtue of its linear dependence, off-axis coma presents less of
challenge for weak lensing than astigmatism.  In Figure~\ref{oaxcoma} we show the
linear growth of off-axis coma with distance from the axis.  The
pattern may be represented by a vector field, directed outward from
the axis.  A coma corrector would produce a vector field directed
inward.  If the corrector is displaced perpendicular to the optical
axis, the difference between the two vectors fields is everywhere a
vector constant.  Therefore two numbers suffice to describe the
mis-correction of off-axis coma.

\subsection{Off-Axis Astigmatism}

Uncorrected, the off-axis astigmatism produced by a paraboloid grows
quadratically with distance from the optical axis.  This is
illustrated in Figure~\ref{oaxastig}, to which some defocus has been added to
produce elliptical point spread functions.

Off-axis astigmatism gets rather little attention because relatively
few telescopes use a field large enough for it to be appreciable.  A
notable exception is the paper by \citet{mcleod95} which describes
turning this aberration to benefit by using it to align a wide field
telescope.  \citet{noethe00} describe a similar test using the
VLT.

We again consider a corrector that has been displaced perpendicular to
the optical axis, so that the wrong correction is applied to the
astigmatic wavefronts.  The mis-corrected images are astigmatic, and
the corresponding point spread functions are shown in Figure~\ref{badastig}.  In
the two panels we have added and subtracted some defocus to produce
elliptical images.  The amplitude of the astigmatism in the miscorrected
images grows {\it linearly} with distance from the center of the
field.

The PSF patterns shown in Figure~\ref{badastig} may look daunting, but it takes
only three numbers to describe them -- the size and direction of the
displacement of the correction from the optical axis and the amount of
defocus.

The linear growth of the size of the mis-corrected astigmatic images
means that the larger the field, the greater the care that must be taken
in aligning the telescope.  An alignment procedure that works
reasonably well for a small field may not work sufficently well for a
larger field.  This is in contrast to the case for coma, where the error
is the same at all field radii.

\subsection{Off-Axis Defocus}

In general the focal surfaces of telescopes are curved rather than flat.
Such ``curvature of field'' may be ``corrected'' either with an optical element
or by the use of a curved detector (or small flat detectors approximating a
curved surface).  A small displacement of the ``corrector'' perpendicular to
the optical axis produces defocus which varies linearly in the direction
in which the corrector was displaced.  The pattern of aberrations
can equivalently be described as a tilt of the focal plane.  Two numbers
suffice to describe this miscorrection of the curvature of field.

Our use of the term ``off-axis defocus'' to describe curvature of field
is non-standard, but it serves to emphasize that three different circularly
symmetric aberration patterns require correction and can be miscorrected.
In \S5, we will use the term tilt instead, which is more intuitive, especially 
when considering how the camera can tilt in response to gravity loading.

\subsection{Miscellany}

Piston errors (i.e. displacements along rather than perpendicular to the optical axis)
of the various correction lenses
will produce under- or over-corrections and leave a residual off-axis
pattern.  
Piston errors will in general produce a constant defocus.
If one piston error is compensated by a second piston error, one
produces higher order spherical aberration which varies as $\rho^4$
in the pupil plane.  

Tilt errors of the lenses produce patterns similar to those produced by translations
perpendicular to the optical axis.

\subsection{Misalignment Summary}

A complete description of the point spread functions due to misaligned
optics requires seven numbers.  Two each for the misalignment of the
coma, astigmatism and defocus corrections perpendicular to the optical
axis and in finaly, the overall defocus of the optical system, which is
produced by a displacement of optics (or the detector) {\it along} the optical axis.

\section{Primary Mirror Deformations}

The primary mirrors of large telescopes deform as a result of thermal
and gravitational stresses.  Not all deformations are equally likely.
A mirror and its support system have ``vibrational'' modes whose
frequencies increase with mirror stiffness.  Static stresses
preferentially deform mirrors in those modes that are least stiff.
The softest modes are almost always: a) a saddle-shaped deformation
that produces wavefront aberrations very much like astigmatism; b) a
bowl shaped mode (more nearly conical when the primary has a
central hole) that produces wavefront aberrations very much like
defocus; and c) a three lobed ``trefoil'' mode ($\cos 3 \theta$ and
$\sin 3 \theta$) \citep{noethe93}.

Large telescopes built after 1990 have ``active'' optics systems that
compensate, in part, for these stresses \citep{noethe93, schechter02}.  
The number of primary mirror modes corrected varies from
telescope to telescope.  In additon such systems typically correct for
translations of the secondary mirror perpendicualar to and along the
optical axis (which produce, respectively, constant defocus and
constant coma).  But not all systems are equally active.  In some
systems only focus is continuously updated in the course of exposures.
Coma and astigmatism are measured at longer intervals.  Lookup tables
are used to correct coma and astigmatism between such measurements.

It is the astigmatism-like and focus-like modes that are of greatest
concern for the measurement of weak lensing.  A small time-invariant defocus
of the telescope will combine with temporal variations in the astigmatism-like
mode to produce elliptical PSFs that are roughly constant across the field.
Temporal variations in the focus-like mode primary mirror will produce ellipticity
if there is any time-invariant astigmatism, perhaps the result
of a small telescope misalignment.  

Most active optics systems do {\it not} correct for temporal variations
in the focus-like primary mirror mode \citep[but see][]{schechter02}. 
But since this mode produces wavefront aberrations that are nearly
degenerate with pistoning of the secondary, most active optics systems
correct for this, at least to first order.\footnote{The difference
between the focus-like primary mirror mode and piston of the secondary
produces spherical-like aberrations.}

\section{Application to Observed PSFs}

\citet{Jarvis04} show ``whisker''
plots showing the shapes of stars observed with the Big Throughput
Camera on the Victor Blanco 4-meter telescope (Figure 1 in their paper).  
They found that the
first principal component of the variation seemed to
correspond to overall defocus.  The plots for images taken very much
inside and very outside focus look like uncorrected or partially
corrected off-axis astigmatism.  Subtracting the best fitting off-axis
astigmatism pattern (with focus varying from image to image) a second
pattern is evident that has elliptical images oriented diagonally in one corner
of one chip and in the perpendicular direction on the opposite chip.
One would get such a pattern if, in addition to the off-axis
astigmatism one had a time-invariant tilt in the focal plane.  This
pattern is also evident in the ``best focus'' frame, for which the
off-axis astigmatism would give no ellipticity if the focal
plane were not tilted.

These results induced us to try to see whether a physical model of the effects of telescope aberrations
might produce a better description of the PSF variation than the purely empirical 
principal component analysis.  

\subsection{The Aberration Model}
\label{model}

The PSF at any location in the image plane is described by an intensity pattern, 
$I(x,y)$.  We always use sky coordinates, measured in arcsec, for $x$ and $y$,
which means that the effects of telescope distortion are implicitly removed.

We start by defining the shape and size of the PSF as the {\it unweighted} second moments of the intensity pattern, $Q$ and $S$:
\begin{align}
Q(x,y) &= \frac{\int dx' dy' I(x',y') \, z'^2}{\int dx' dy' I(x',y')} \\
S(x,y) &= \frac{\int dx' dy' I(x',y') \, |z'|^2}{\int dx' dy' I(x',y')}
\end{align}
where $z = x + i y$ gives the location in the image plane, and $x'$
and $y'$ are integrated over the extent of the light distribution for  
a single PSF (i.e. a single star) centered at $z$.

The essential causes of the PSF shape and size are the telescope
optics and the atmosphere.
There are some other minor causes, like diffusion in the CCD, but we neglect these in this model.
We assume that the effects are separable, so $Q$ and $S$ are sums of optics effects and
atmospheric effects.  

%As described in \S ?? above, 
For the optics contribution we can take the photons entering the
telescope to be 
exactly parallel.  We also assume that the density of photons is
uniform across the pupil.  Then the intensity as a function 
of position in the field, $I(x,y)$ can be obtained by a)
subdividing the pupil into (infinitesimal) patches of equal area, b) taking the
gradient of the wavefront at each patch, c) taking that gradient to be
proportional to the displacement of the light from the nominal image
position\footnote{
The constant of proportionality is related to the focal length of the telescope.
In practice this constant is absorbed into the aberration coefficients, so it is
safely neglected.} 
and d) summing the contributions to $I(x,y)$ from all patches.

Taking $u$ and $v$ to be the coordinates on the pupil, the PSF at a
point in the image plane is given by the pupil plane integrals
\begin{align}
Q(x,y) &= \frac{1}{\int du dv} \int du dv \left(\frac{\partial\delta \lambda}{\partial u} + i \frac{\partial \delta \lambda}{\partial v}\right)^2 \\
S(x,y) &= \frac{1}{\int du dv} \int du dv \left|\frac{\partial\delta \lambda}{\partial u} + i \frac{\partial \delta \lambda}{\partial v}\right|^2 
\end{align}
%Note that
%these depend {\it quadratically} on the gradient of the wavefront.
%Other contributions to the moments can arise either upstream and
%downstream of the telescope (e.g. atmospheric seeing and charge
%diffusion).  These contributions are thought to contribute {\it
%linearly} to the unweighted moments $Q$ and $S$.  
We insert the wavefront errors due to defocus, astigmatism, coma,
and spherical aberration from 
Equations~\ref{del_defocus}, \ref{del_astig1}, \ref{del_astig2},
\ref{del_coma1}, \ref{del_coma2}, and \ref{del_sph} and define
\begin{align}
d &= A_{\rm defocus} \nonumber \\
a &= (A_{\rm astig-c} + i A_{\rm astig-s})\nonumber  \\
c &= (A_{\rm coma-c} + i A_{\rm coma-s})\nonumber  \\
s &= A_{\rm sph}
\end{align}
to get
\begin{align}
Q(x,y) &= \frac{1}{\int du dv} \int du dv
\left( 2d w + 2a w^* + c(2 |w|^2-1) + c^*w^2 + 4 s |w|^2 w\right)^2 \\
S(x,y) &= \frac{1}{\int du dv} \int du dv
\left| 2d w + 2a w^* + c(2 |w|^2-1) + c^*w^2 + 4 s |w|^2 w\right|^2 
\end{align}
where $w = u+i v$ and ${}^*$ denotes complex conjugate.
Integrating over the circular pupil, this simplifies to:
\begin{align}
\label{q1}
Q(x,y) &= 4 \left(d + \frac{4}{3} s\right) a + \frac{1}{3} c^2 \\
\label{s1}
S(x,y) &= 2 \left(d + \frac{4}{3} s\right)^2 + 2 |a|^2 + \frac{2}{3} |c|^2 + \frac{4}{9}s^2 
\end{align}

For the defocus value, we include four effects given by four
parameters:
$d_0$ is the overall defocus of the camera.
$d_1$ is a tilt of the wavefront as it strikes the camera,
which leads to the defocus increasing linearly with
field position because the divergence of the rays is proportional to
the distance from the center.  Thus, the value at any given
position is the ``dot product'' of $d_1$ with (x,y).  In complex
notation this is $\Re(d_1 z^*)$. 
In addition, we allow for a separate defocus and tilt of each chip in
the camera, which we call $d_{0,\rm chip}$ and $d_{1,\rm chip}$.
\begin{equation}
\label{d_z}
d(z) = (d_0+d_{0,\rm chip}) + \Re \left((d_1+d_{1,\rm chip})
\frac{z^*}{R}\right) 
\end{equation}
$R$ is the radius of the field of view, which is included so that
$d_0$ and $d_1$ have the same units, arc seconds. 

The astigmatism value in our model has two components: The primary 
astigmatism is $a_0$, which is constant across the field of view.  The
miscorrection of the off-axis astigmatism 
is $a_1$.  The effect of the $a_1$ term is derived from taking two
copies of the PSF pattern in  
Figure~\ref{oaxastig}, shifting them and taking the difference.  To first order,
this is just the derivative.  So, since the astigmatism
in Figure~\ref{oaxastig} is proportional to $z^2$, 
the off-axis astigmatism represented by $a_1$ is proportional to $z$.
\begin{equation}
\label{a_z}
a(z) = a_0 + a_1 z
\end{equation}
Both $a_0$ and $a_1$ have units of arc seconds.

As discussed in \S\ref{off-axis_coma}, the net effect of coma in the presence of a 
possibly misaligned corrector lens is a single value, $c(z) = c_0$.  
Since the misalignment may vary from one exposure to the next,
we let $c_0$ be different for each exposure.
There are generally no
other significant sources of coma, so this is the only one we include in our model.

We do not expect spherical aberration to vary from one exposure to another,
but we do allow for the possibility of some in the telescope design or due to
possible slight misplacements of the optical elements.  Thus, we could expect there to be 
a single value of $s(z) = s_0$ for all exposures.  However, such a value is completely degenerate with
the overall defocus, $d_0$, and the atmospheric seeing (discussed below), because only the 
combination $d+(4/3) s$ appears in the above formulae.  Therefore, this combination is taken to
be the effective defocus value in the model, and we can thus ignore $s$ entirely.

In addition to the optical model given above, we also add the seeing, $S_{\rm atm}$ to
the size.  The seeing is not an optical effect of the telescope, but it is obviously the dominant contributor
to the PSF size.  Therefore, it is absolutely essential to the model.  This term also absorbs the 
$(4/9) s^2$ term that is neglected by ignoring spherical aberration.  

The effects of guiding and the average anisotropy of the atmosphere are generally also important.
However, since we already have the $c_0$ term in the formula for $Q$, these effects are 
subsumed into that value.  There is no way to distinguish coma from guiding errors using
only the second moments of the PSF.  The trefoil aberration we mentioned above also leads to a
constant term in $Q$ (through interactions with the coma), but this is similarly indistinguishable
from coma.

We also found it necessary to include an additional function for each
of $Q$ and $S$ to describe the base {\it static} pattern 
to which the effects of the aberrations for each exposure 
are added.  These functions are called $Q_{\rm static}$ and $S_{\rm static}$.  
We use a simple 5th order polynomial for each of these.  

So we have the following model for the PSF shape and size as
a function of position in the focal plane:
\begin{align}
Q &\approx Q_{\rm model}(x,y) \equiv 
4 d(z) a(z) + \frac{1}{3} c_0^2 + Q_{\rm static}(x,y) \\
S &\approx S_{\rm model}(x,y) \equiv
2 d(z)^2 + 2 |a(z)|^2 + \frac{2}{3} |c_0|^2 +  S_{\rm atm} + S_{\rm static}(x,y)
\end{align}
where $d(z)$ and $a(z)$ are given above in equations~\ref{d_z} and \ref{a_z}.

It is important to note that the terms due to optical
aberrations enter the moment equations {\it quadratically}.  As the
moment model is linear in $Q_{\rm static}$ and $S_{\rm static}$, 
the implication is that these do {\it not} arise from distortions of
the wavefront.  This raises the question of just what the source of
these static terms might be or if in fact this is the correct functional form to use.

With real data, unweighted moments are not feasible.  Instead, 
we define the shape $Q$ and size $S$ of each observed star with
Gaussian weights: 
\begin{align}
Q(x,y) &= \frac{ \int dx'  dy' ~ z'^2 e^{-|z'|^2/2\alpha^2} I(x',y')
}{ \int dx' dy' ~ e^{-|z'|^2/2\alpha^2} I(x',y')} \\ 
S(x,y) &= \frac{ \int dx'  dy' ~ |z'|^2 e^{-|z'|^2/2\alpha^2} I(x',y')
}{ \int dx' dy' ~ e^{-|z'|^2/2\alpha^2} I(x',y')}  
\end{align}
where the scale size, $\alpha$, of the Gaussian
is chosen such that $S = \alpha^2$. (We iterate the choice of $\alpha$ until this is true.)  $Q$ and $S$
both have units of arc seconds squared.

Another important consideration is the orientation of the coordinate 
system we use to 
define $z$.  The two natural choices are equatorial coordinates, which are oriented east/west and north/south, and alt-az coordinates (also known as ``horizon'' coordinate), where $y$ is oriented so that it corresponds to 
the direction of gravity, or up/down, and $x$ is then left/right.  Different physical causes for 
the parameters are more natural in one coordinate system or the other.  Anything that is fixed
with respect to the orientation of the camera is more natural in equatorial coordinates.  Anything
that has to do with the gravity loading of the telescope is more natural in alt-az coordinates.

Anticipating that some of the parameters may be connected with gravity loading, 
we adopt the alt-az choice for $z$ in $d(z)$ and $a(z)$\footnote{
Since the coordinate systems differ only in a phase factor ${\rm exp}(i \phi)$, this choice
does not affect the $\chi^2$ of the fit, only the covariance matrix and the correlations with 
hour angle, declination, etc.}.
However, for the static 
functions $Q_{\rm static}(x,y)$ and $S_{\rm static}(x,y)$, we found that it was better to
keep $x$ and $y$
in the chip-oriented equatorial coordinates.  Similarly, the chip tilts 
obviously need to be in equatorial coordinates as well, since the direction of the tilt moves with
the camera.  

There are a few approximations implicit in this model formula, which
are worth articulating. First, the formulae for $Q$ and $S$ given in equations~\ref{q1}
and \ref{s1} are
derived for unweighted moments. However, unweighted moments are
impossible on real data, so we use the above Gaussian weight, 
but this means that the formulae are no longer quite
correct. 
%{\sc Might the correct thing be a Gaussian weight in eqns 14,15?  Since the Fourier transform 
%of a Gaussian is a Gaussian...}

Second, the single value for $S_{\rm atm}$ and the incorporation of
the anisotropy of the atmosphere into the single value $c_0^2$ ignore the real
variation of the atmospheric component across the field of view.  This
should not be a significant problem, since the discrepancy for each exposure
will be a stochastic function, which will average out over very many
exposures (see \citealp{Jain06} for applications to
shear correlation measurements).  
%The neglected spatial variation of 
%the atmospheric contribution will likely dominate the 
%residual PSF patterns when we fit our model to the data. 

Third, we have measured the distortion from the actual positions of
stars and galaxies, using 
overlapping images as well as the USNO astrometric catalog to
calculate the overall astrometric solution. 
However, this solution is constrained to be a 5th order function that
is the same for all images 
plus a simple translation and rotation for each exposure.  If the telescope distortion varies
from one exposure to the next (for example, from the same kind of decollimations that led to 
the tilt and off-axis astigmatism)  then this will lead to errors.

Fourth, the derivation of equations~\ref{q1} and \ref{s1} assumed uniform illumination
across the pupil.  This is not actually true.  The prime focus camera obscures the center
of the pupil.  And near the edges of the image, vignetting may be in issue as well.

Finally, there are other higher order aberrations than the ones we have included,
which are generally only referred to
by their corresponding Zernike polynomial.  Similarly, while our expressions for 
off-axis focus, astigmatism, and coma are correct to first order, there may be terms 
at higher order in $x$ and $y$ that we have neglected.

\subsection{Results}
\label{results}

%\subsection{Blanco telescope images with BTC and Mosaic cameras}

We applied the above methodology for the telescope aberrations to 
data taken using the 
Mosaic camera of the Blaco 4-meter telescope at Cerro Tololo Interamerican Observatory
(CTIO).  The camera has 8 chips, each 2048 $\times$ 4096 pixels with 0.27 arc-second 
pixels.  The field of view has a radius ($R$ in the formulae above) of 18 arcminutes.

We analyzed the PSF shapes from 383 exposures taken in January and July, 2000.
For each unsaturated star brighter than 21st magnitude, we measured $Q$ and $S$ and
fit them to the above model.

In total, the model has 10 real-valued numbers for each exposure, of which
9 are mis-alignments ($d_0, d_1, a_0, a_1, c_0$) and the last is seeing ($S_{\rm
atm}$), plus 82 numbers that 
are constant for all exposures (see Table 1 for the full list of parameters).  
For 383 exposures, this comes to over 3900 numbers  
being fit for.  To constrain these values, we have almost 400,000 stars, each with three numbers
to use, so the model is safely over-constrained.

It would be completely impractical to fit all of these parameters at once, especially since the 
model is non-linear.  So we instead alternate fitting the parameters for each
exposure and then the static telescope parameters until the fit converges.

\begin{deluxetable}{ccccccc}
\tablewidth{0pt}
\tablecaption{Parameters used in the PSF Model}
\tablecolumns{7}
\tablehead{

\colhead{Parameter} &
\colhead{Description} &
\colhead{Number of Real} &
\colhead{Average} &
\colhead{Standard} &
\multicolumn{2}{c}{Marginal Utility} \\

\colhead{} &
\colhead{} &
\colhead{Components} &
\colhead{Value} &
\colhead{Deviation} &
\colhead{~~~$\chi^2_Q$~~~} & \colhead{~~~$\chi^2_S$~~~}
}

\startdata

\sidehead{\it Exposure-specific parameters:}

$d_0$ & Primary defocus & $1 \times 383$ & $-0.006$ & $0.053$ & -0.4 & 3.2  \\
$d_1$ & Off-axis defocus & $2 \times 383$ & $\begin{array}{c}0.009 \\ -0.003\end{array}$ & 
$\begin{array}{c}0.038\\0.040\end{array}$ & 3.7 & 4.2  \\
$a_0$ & Primary astigmatism & $2 \times 383$ & $\begin{array}{c}0.014\\-0.011\end{array}$ &
$\begin{array}{c}0.099\\0.089\end{array}$ &  7.2 & -0.5 \\
$a_1$ & Off-axis astigmatism & $2 \times 383$ & $\begin{array}{c}0.001\\-0.002\end{array}$ &
$\begin{array}{c}0.066\\0.055\end{array}$ & 1.8 & 0.7 \\
$c_0$ & Coma/guiding shape & $2 \times 383$ & $\begin{array}{c}-0.039\\-0.010\end{array}$ & 
$\begin{array}{c}0.109\\0.124\end{array}$ & 4.7 & 1.1 \\
$S_{\rm atm}$ & Seeing size & $1 \times 383$ & $0.348$ & $0.138$ & 5.7 & 2.7 \\

\sidehead{\it Static telescope parameters} 

$d_{0,\rm chip}$ & Chip height offsets & 7 & & & 0.3 & 1.5 \\
$d_{1,\rm chip}$ & Chip tilts & 14 & & & 0.4 & 0.7  \\
$Q_{\rm static}$ & Static shape pattern & 40 & & & 1.7 & 0.3  \\
$S_{\rm static}$ & Static size pattern & 20 & & & -0.1 & 2.0  

\enddata

\label{paramstable}

\end{deluxetable}

Table~\ref{paramstable} lists each of the parameters, along with a summary description and
the number of (real) values being fit for.  It also lists the average values of each of
the parameters that change for each exposure, along with their standard deviation.  
The last column, Marginal Utility, is discussed below in \S\ref{results}.

\begin{deluxetable}{rcccccccccc}
\tablewidth{0pt}
\tablecaption{Covariance Matrix of Exposure Parameters}
\tablecolumns{11}
\tablehead{ ~& ~& ~& ~& ~& ~& ~& ~& ~& ~& ~}

\startdata

~ & $d_0$ & \multicolumn{2}{c}{$d_1$} & \multicolumn{2}{c}{$a_0$} &
\multicolumn{2}{c}{$a_1$} & \multicolumn{2}{c}{$c_0$} & $S_{\rm atm}$ \\
$d_0$ & $1$ & $0.15$ & $0.16$ & $0.09$ & $-0.07$ & $0.01$ & $0.18$ & $0.04$ & $-0.03$ & $-0.09$ \\
$d_1$ & 
  $\begin{array}{c} 0.15 \\ 0.16 \end{array}$ & $\begin{array}{c} 1 \\ 0.36 \end{array}$ &
  $\begin{array}{c} 0.36 \\ 1 \end{array}$ & $\begin{array}{c} 0.23 \\ -0.04 \end{array}$ &
  $\begin{array}{c} -0.01 \\ 0.09 \end{array}$ & $\begin{array}{c} 0.75 \\ 0.43 \end{array}$ &
  $\begin{array}{c} 0.29 \\ 0.73 \end{array}$ & $\begin{array}{c} -0.15 \\ -0.14 \end{array}$ &
  $\begin{array}{c} 0.22 \\ -0.13 \end{array}$ & $\begin{array}{c} -0.14 \\ 0.01 \end{array}$  \\
$a_0$ & 
  $\begin{array}{c} 0.09 \\ -0.07 \end{array}$ & $\begin{array}{c} 0.23 \\ -0.01 \end{array}$ &
  $\begin{array}{c} -0.04 \\ 0.09 \end{array}$ & $\begin{array}{c} 1 \\ 0.29 \end{array}$ &
  $\begin{array}{c} 0.29 \\ 1 \end{array}$ & $\begin{array}{c} 0.20 \\ 0.02 \end{array}$ &
  $\begin{array}{c} -0.11 \\ 0.00 \end{array}$ & $\begin{array}{c} -0.04 \\ 0.07 \end{array}$ &
  $\begin{array}{c} 0.01 \\ -0.20 \end{array}$ & $\begin{array}{c} -0.18 \\ 0.19 \end{array}$  \\
$a_1$ & 
  $\begin{array}{c} 0.01 \\ 0.18 \end{array}$ & $\begin{array}{c} 0.75 \\ 0.29 \end{array}$ &
  $\begin{array}{c} 0.43 \\ 0.73 \end{array}$ & $\begin{array}{c} 0.20 \\ -0.11 \end{array}$ &
  $\begin{array}{c} 0.02 \\ 0.00 \end{array}$ & $\begin{array}{c} 1 \\ 0.48 \end{array}$ &
  $\begin{array}{c} 0.48 \\ 1 \end{array}$ & $\begin{array}{c} -0.10 \\ -0.15 \end{array}$ &
  $\begin{array}{c} 0.19 \\ 0.02 \end{array}$ & $\begin{array}{c} -0.09 \\ 0.10 \end{array}$  \\
$c_0$ & 
  $\begin{array}{c} 0.04 \\ -0.03 \end{array}$ & $\begin{array}{c} -0.15 \\ 0.22 \end{array}$ &
  $\begin{array}{c} -0.14 \\ -0.13 \end{array}$ & $\begin{array}{c} -0.04 \\ 0.01 \end{array}$ &
  $\begin{array}{c} 0.07 \\ -0.20 \end{array}$ & $\begin{array}{c} -0.10 \\ 0.19 \end{array}$ &
  $\begin{array}{c} -0.15 \\ 0.02 \end{array}$ & $\begin{array}{c} 1 \\ -0.05 \end{array}$ &
  $\begin{array}{c} -0.05 \\ 1 \end{array}$ & $\begin{array}{c} 0.13 \\ -0.08 \end{array}$  \\
$S_{\rm atm}$ & $-0.09$ & $-0.14$ & $0.01$ & $-0.18$ & $0.19$ & $-0.09$ & $0.10$ & $0.13$ & $-0.08$ & $1$ \\ 
\\
\tableline
HA & $-0.03$ & $0.28$ & $-0.06$ & $-0.01$ & $-0.06$ & $0.18$ & $0.07$ & $-0.06$ & $0.25$ & $-0.03$ \\
Dec & $0.09$ & $0.37$ & $0.39$ & $0.06$ & $0.05$ & $0.27$ & $0.37$ & $-0.25$ & $0.09$ & $0.00 $ \\
ZD & $0.13$ & $0.20$ & $0.28$ & $0.07$ & $0.03$ & $0.11$ & $0.19$ & $-0.25$ & $-0.05$ & $0.03 $

\enddata

\label{covartable}

\end{deluxetable}

Table~\ref{covartable} gives the normalized covariance matrix for the six exposure-specific
parameters.  The values are normalized by the standard deviations so the diagonal 
elements are unity.  The matrix is $10 \times 10$, rather than $6 \times 6$, since each complex
quantity is treated as two separate real values for this purpose.
The table also includes the correlations with hour angle, declination, and zenith distance to 
indicate which parameters are connected with gravity.  
%See \S\ref{gravity} for more about this.

The largest correlation apparent in the data is $d_1$ with $a_1$, both of which also correlate
with the direction of gravity, namely the declination and zenith distance.

\begin{figure}[t]
\epsscale{1.0}
\plottwo{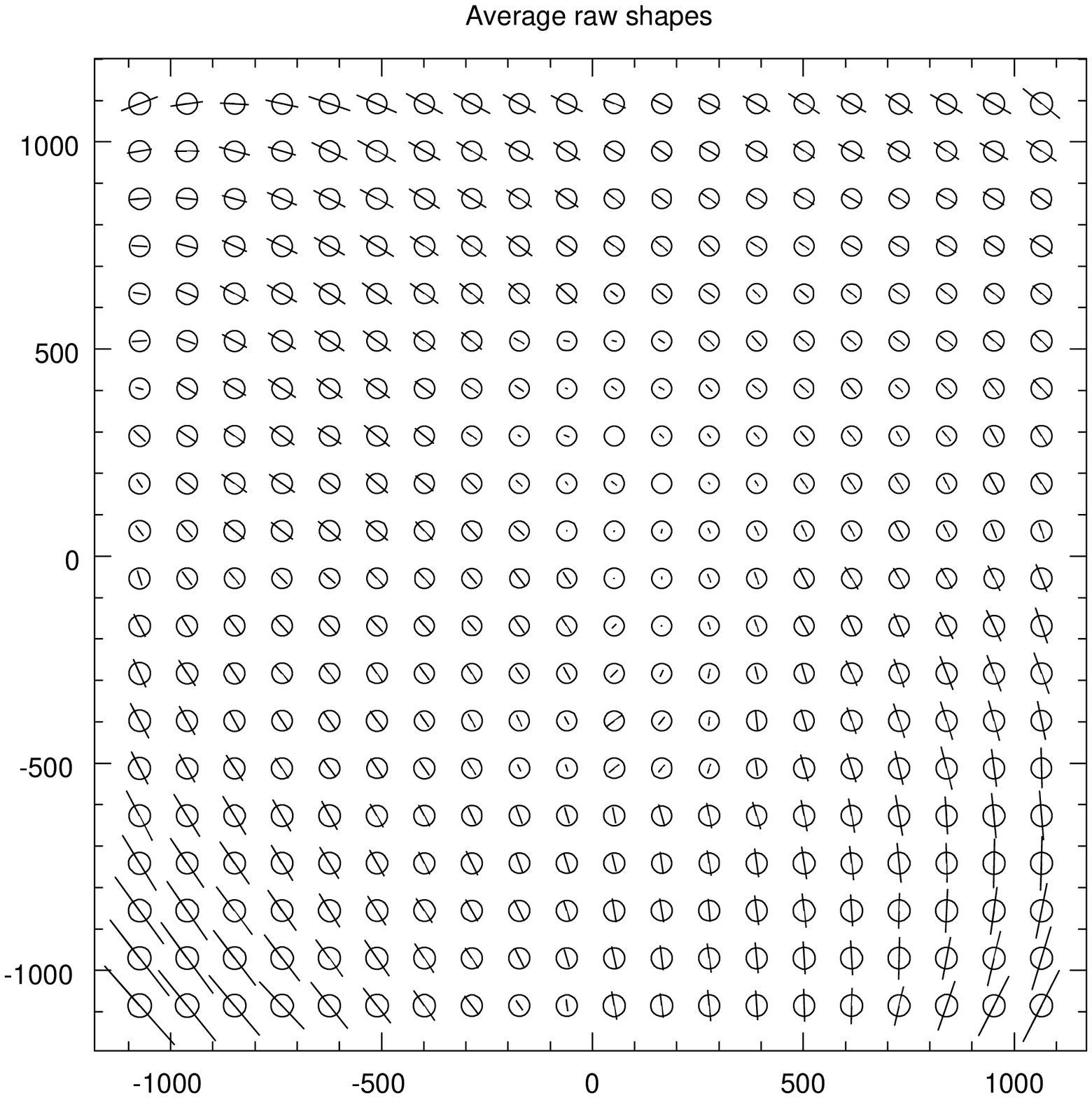}{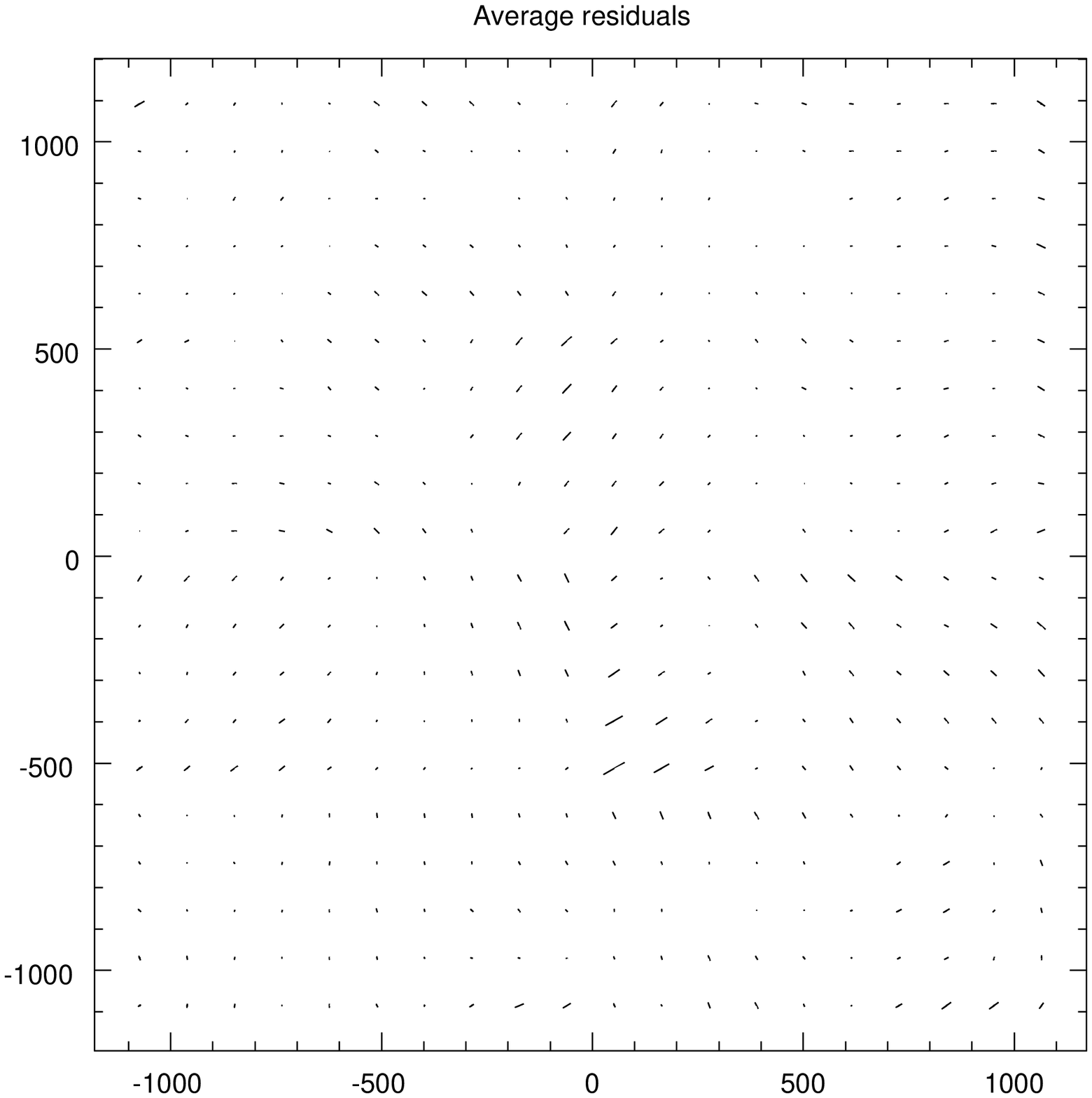}
\caption[]{\small The left panel shows the average PSF pattern.  The ``whisker'' lines 
correspond to the average $Q$ values.  The direction indicates the orientation of the 
ellipse, and the length is proportional to $|Q|$.  The circles correspond to the 
average $S$ values, the size being proportional to the size of the
PSF.  The right plot shows the difference between the measured and fitted PSF values.}
\label{raw}
\end{figure}

Figure~\ref{raw} is a ``whisker plot'' of the mean PSF
pattern averaged over all 383 exposures.  The lines (``whiskers'')
indicate the shape, $Q$, at each location.  The orientation 
of the line is the same orientation as the ellipse described by $Q$, and the length
of the line is proportional to $|Q|$.  
The circles are used to represent the size, $S$, at each location. The size of the circle
is proportional to the value of $S$.

We can start to see how well the model fits this pattern by plotting the average of the
model $Q$ and $S$ values rather than the actual measured values and look at the residual
(right panel of Figure~\ref{raw}).  The aberration model apparently does manage to describe the 
data fairly well; however, this is not necessarily a very good test, since we are mostly seeing the static
pattern here, which is directly fit using a fifth order polynomial. 

A better test is to look at the reduced $\chi^2$ value of the residual PSF anisotropy:
\begin{align}
\chi^2_Q &= \frac{1}{(N_{\rm star}-N_{\rm param})} \sum_i
\frac{|Q_i-Q_{\rm model}|^2}{\sigma_{Q,i}^2}  \\
\chi^2_S &= \frac{1}{(N_{\rm star}-N_{\rm param})} \sum_i
\frac{(S_i-S_{\rm model})^2}{\sigma_{S,i}^2} 
\end{align}
The total relevant $\chi^2$ value would be $\chi^2_Q + \chi^2_S$, but we find it 
instructive to split the quantity into the two parts.

We find the values $\chi^2_Q = 7.5$ and $\chi^2_S = 7.0$, compared to initial 
values of 90.6 and 8014 (i.e. where $Q_{\rm model} = S_{\rm model} = 0$).  So,
we could say that the model is accounting for over $90\%$ of the full
description of the PSF shape and over $99\%$ of the description of the size.
(The initial size chisq is dominated by the seeing, so the majority of this latter decrease
is not an impressive test of the model.)

It is interesting to examine how important each
of the parameters used in the model is to the final solution.  In particular, we find the
marginal utility of each parameter for each of the $\chi^2$ values, which  
means the amount by which $\chi^2$ increases when each parameter in turn is removed
from the model, allowing the other parameter values to adjust to the new model. 
The primary astigmatism is found to be the
most important for the shape, causing an increase of 7.2 when this parameter
was removed from the model. (That is, $\chi^2_Q = 14.7$ in that case.)
The marginal utilities of each parameter is listed in Table~\ref{paramstable}\footnote{
Some of these marginal utilities are negative indicating that the fit became slightly better for
either $Q$ or $S$.
The maximum likelihood solution minimizes the sum of $\chi_Q^2$ and $\chi_S^2$, which
always increases when a parameter is removed.}

The only parameter besides the seeing with a significantly non-zero average 
value is the off-axis astigmatism.  This is interesting, since we expect this parameter to be telling
us about displacement of optical elements like corrector lenses from the optical axis
(see Figure~\ref{badastig}).  So this
may indicate that there is an optical element that is permanently misaligned.  
However, since the constant offset is in alt-az coordinates, the ``permanent'' misalignment
does move as the telescope changes orientation with respect to the horizon.

\section{Discussion}

We have presented low order telescope image aberrations as relevant
for weak lensing studies with wide field imaging data. We have found
that a fairly simple model of the PSF variation, based on an
understanding of how focus and astigmatism work and how they can vary
across the field, is able to describe approximately 90\% of the
variation of PSF anisotropy seen in real data. Details of our model
and results are given in Sections 5.1 and 5.2. 

We find that the astigmatism parameters in our model are
moderately correlated with the direction of gravity, suggesting that gravity loading is
to some extent responsible for the telescope aberrations. 
%Figures~\ref{d1_plot}, \ref{a0_plot} and \ref{a0_plot2}
%show these correlations; they are discussed in \S\ref{gravity}, as
%are other features that we do not understand. 

At least a portion of the residual PSF is expected to be due to the atmospheric
seeing varying across 
the field of view in each exposure. Newer telescope are expected to
have better optical performance; however, even with such improvements,
our results suggest that the contribution of telescope optics will remain
a significant contributor to PSF anisotropy. This is especially true
for surveys that use large numbers of exposures for every field to reduce the
atmospheric contribution, since the atmospheric contribution is
stochastic in time while the contribution of the telescope optics in consort 
with gravity loading is not.

Our success in modeling PSF patterns in real data is encouraging for
planned surveys that aim to optimize their image quality for weak
lensing.  On one hand, one can hope to use data taken over the
course of the survey to to correct telescope misalignments as they
occur.  On the other, to the extent that such misalignments are not
corrected, measurements of the misalignments permit compensation in
the data taken using a physical model.

These points have recently been made by 
\citet{Ma08}, who use ray tracing to map the PSF moments
produced by displacements and tilts of the secondary of the
space-based SNAP telescope.  Their approach differs from the one taken
here primarily in that they use detailed models for each of the
misaligned elements while we assume that {\it all} misalignments
produce astigmatism, defocus and coma patterns that differ {\it only}
in orientation and magnitude.  Secondarily, they fit third moments as
well as second moments, giving them a better handle on coma and breaking
the degeneracy between translations and rotations of their secondary
mirror.  
%Were we to try, as they do, to attribute our misalignments to
%particular optical elements we might run into the same uncertainties
%they do.

More work needs
to be done to improve the model and apply it to both understand
telescope performance and to gain better precision in PSF
interpolation. 
In \S\ref{model}, we listed a number of possible shortcomings
of our aberration model.  It is unclear to us which of these is the
most important to address to make further improvements.  
We suspect that accounting for the weighted moments may be
the most promising direction to proceed. 

Another hint is the demarcation of chip boundaries in the residual PSF pattern
in Figure~\ref{raw}.
Perhaps there is some possible improvement from better modeling of
the chips: for example, using a second or third order function for each chip.
Also, while we expect that our off-axis astigmatism term is the most important
component of astigmatism after the constant primary mirror 
astigmatism, it is possible
that there are other functional forms which are also important to include.

It is worth comparing the efficacy of this model with the purely
empirical PCA description of \citet{Jarvis04}.  Since the model uses
10 real numbers for each exposure, we compare it to the PCA
model with 5 principal components, each with a complex coefficient. 
The PCA model leaves a residual  $\chi_Q^2+\chi_S^2$ of $8.5+7.0$, compared to the
aberration model's $7.5+7.0$.  So the aberration model is only doing slightly
better than the purely empirical method (though a
fair comparison is admittedly difficult to formulate). 
This conclusion may differ for a different telescope or for a
larger dataset as expected from upcoming surveys. We leave a more
detailed analysis of the performance for follow-up work. 

Finally, note that a physical model is unlikely to be able to account for
all of the observed PSF pattern.  Even aside from atmospheric refraction, 
the effect of resonance shaking of the
telescope, for example, would be exceedingly difficult to model 
correctly.  \citet{Jarvis04} found
one of their principal components to correlate with the wind speed in a particular 
direction.  The most likely explanation seems to be that the wind was able to excite
a resonance mode in the telescope, which would correspond to an integral of several
different misalignment effects coupled in a complicated way.  It is hard to imagine that 
this kind of effect could be effectively included in a physical model like the 
one we describe here.  Therefore, any practical use of this technique
would probably have to be followed by a principal component analysis to remove the various
more complicated effects that are not included in the model.

%In conclusion, the methods presented here are a promising start to a
%more accurate description of the variation of the PSF.  More work needs
%to be done to improve the model and apply it to both understanding
%telescope performance and to gain better precision in PSF
%interpolation. 

\acknowledgements
We thank Gary Bernstein, Steve Kent and Andy Rasmussen for helpful
discussions.  This work is supported in part by NSF grant
AST-0607667 (at Penn), AST-0602010 (at MIT), the 
Department of Energy and the Research Corporation.

\end{document}